# Identification of defects responsible for optically detected magnetic resonance in hexagonal boron nitride

A. Sajid[1,2,3], Kristian S. Thygesen[3], Jeffrey R. Reimers[1,4], and Michael J. Ford[1]

[1]University of Technology Sydney, School of Mathematical and Physical Sciences, Ultimo, New South Wales 2007, Australia.

[2]Department of Physics, GC University Faisalabad, Allama Iqbal Road, 38000 Faisalabad, Pakistan

[3]CAMD, Department of Physics, Technical University of Denmark, 2800 Kgs. Lyngby, Denmark.

[4]International Centre for Quantum and Molecular Structures and Department of Physics, Shanghai University, Shanghai 200444, China.

Email sajal@dtu.dk; thygesen@fysik.dtu.dk; Jeffrey.reimers@uts.edu.au;

mike.ford@uts.edu.au

**Abstract:** Crystal defects in the two-dimensional insulator hexagonal boron nitride (hBN) can host localised electronic states that are candidates for applications in quantum technology, yet the precise chemical and structural nature of the defects measured in experiments remains unknown. Recently, optically detected magnetic resonance (ODMR) has been measured from defects in hBN for the first time, providing rich information. In one case a tentative assignment has been made to the $V_B^-$ defect. Here, we report density-functional theory (DFT) calculations of the energetics, zero field splitting parameter, $D$, and hyperfine coupling tensor, $A$, that confirm this assignment. We also show that a second ODMR measurement is consistent with the $V_N^-$ defect. In both cases, the location of the defect at the edge or centre of layers is important. This advances our atomic-scale understanding of crystal defects in hBN, which is necessary for harnessing the potential of these systems for quantum technology applications.

The discovery[1,2,3,4] in 2016 of single-photon emission (SPE) from defects in hexagonal boron nitride (h-BN) has inspired significant experimental and theoretical research in an effort to understand the chemical origin of this emission. Excitement stems from the possibility that such SPE could be harnessed by nanophotonic[5,6,7,8] industries to deliver applications. Their exploitation, however, requires understanding of the chemical nature of the colour centres in h-BN and the ability to control and tune their structural arrangements and spectroscopic properties. While various colour centres have been proposed as potential sources of SPEs in h-BN,[9,10,11,12] none have been firmly identified.[10] In parallel to this work, defect centres in h-BN have also been demonstrated to produce strong paramagnetic effects, and for these chemical assignments have indeed been developed.[10]

A dramatic recent advance is the observation of room-temperature optically detected magnetic resonance (ODMR),[13,14] a phenomenon involving both measurement of ground-state magnetic properties and photoluminescence (PL) on the same defect. This not only makes it much easier to identify the chemical nature of a colour centre, but also suggests possible future applications in quantum information and other technologies. In one ODMR report, by Chejanovsky et al.,[14] the ZFS was too small to measure (magnitude < 4 MHz) and, the hfc was determined to be around 10 MHz with an angular dependence consistent with an unpaired electron in a $\pi$ orbital, tentatively associated with a substitutional carbon impurity or other low mass atom. Instead, we suggest that this could be associated with the $V_N^-$ defect involving a negatively charged nitrogen vacancy located near the edge of a h-BN layer. In the other report, by Gottscholl et al.,[13] a ZFS of -3.5 GHz was deduced with a triplet defect ground state, notwithstanding some ambiguity as to the sign of this quantity. This latter work provides an unambiguous determination of the defect ground-state spin parameters. We use DFT calculations to interpret these results and demonstrate that the defect responsible is indeed the negatively charged boron vacancy $V_B^-$. If the sign of the ZFS is negative, then we identify the defect at a layer edge, whereas if the sign is positive, then we identify the defect to be embedded in the bulk region of the layer.

DFT calculations may be applied to predict the spectroscopy and magnetic properties of defects.[10] The prediction of optical spectroscopic properties, such as the zero phonon line (ZPL) energy, by DFT is very difficult as most defect states are open-shell in nature, displaying multi-reference characteristics, whereas DFT is intrinsically a single-reference approach.[15] Nevertheless, time-dependent DFT (TDDFT) can be reliable for defect excited states, provided that it is performed based on a well-represented reference state.[10] Also, DFT is typically quite accurate when it comes to the prediction of ground-state magnetic properties.[10] Indeed, DFT predictions of the HF and ZFS properties of the $N_V^-$ centre in diamond[16, 17] were important for the identification of its chemical structure. Further, previously observed paramagnetic signals from h-BN defects have also been assigned by DFT as the neutral nitrogen ($V_N$) and boron ($V_B$) vacancies.[18, 19] We therefore expect that DFT calculations will be reliable for the identification of defects associated with the observed ODMR.

Similar calculations pertaining to the Gottscholl et al.[13] observations have just appeared in a preprint[20] that also assigns its ODMR to bulk $V_B^-$ defects, assuming that the original publication incorrectly reported the sign of the ZFS. They showed that distortions from the intrinsic $D_{3h}$ symmetry of a bulk defect are required to explain the observed non-zero value for the off-axial ZFS parameter of *E* = 50 MHz. Their DFT calculations of the excited-state spectroscopy of this defect are also consistent with the experimentally observed photoluminescence, notwithstanding the issues associated with this approach. We consider the location of the defect at the edge as an alternative distortion mechanism, and the effect of surrounding bulk layers on the layer containing the defect.

A difficult, but important, aspect of the use of calculations to assign defects is that all plausible defects need to be considered and differentiated between. We consider not only the $V_B^-$ and $V_N^-$ defects as possible causes of the two sets of observed ODMR, but also a defect involving an internal boron dangling bond[11] and the $C_N^{\pm}$ defects in which either a positively or negatively charged carbon replaces a nitrogen, akin to suggestions made by Chejanovsky et al.[14] concerning the possible nature of the defect they observed. Out of all the considered defects, only $V_B^-$ or $V_N^-$ show spin structures consistent with the experimental results.

SPEs in h-BN have been categorized into two broad groupings:[2, 10] Group-1 emitters ZPL energies of 1.8-2.2 eV, total Huang-Rhys factors of 0.9-1.9, and emission reorganization energies (half bandwidths) of 0.06-0.16 eV, whereas Group-2 emitters have ZPL energies of 1.4-1.8 eV, total Huang-Rhys factors of 0.25-0.6, and emission reorganization energies of 0.05-0.08 eV. The ODMR emitters observed by Chejanovsky et al.[14] are fully characteristic of Group-2. In contrast, the ODMR signal observed by Gottscholl et al.[13] originates from an ensemble of defects, with single-defect measurements not currently possible owing to weak

PL intensity. It is therefore difficult to determine whether the observed PL line shape is characteristic of the defect itself or else dominated by thermal or inhomogeneous broadening. The observed emission energies near 1.5 eV are in the range typical of Group-2 emitters. These emitters[13] could be examples of a new group, however, with an important difference being that Group-1 and Group-2 emitters are very bright (internally varying by ca. only an order of magnitude in luminescence), whereas the emitters probed in[13] are very weak.

Much work has considered the location of defects within h-BN samples.[10] Of note is that Group-2 emitters have been observed specifically at or near grain boundaries and flake edges,[21, 22] as well as near structural defects in the basal plane of the h-BN.[23] In addition, they have also been observed well away from edges and grain boundaries.[24, 25] It is therefore important that experimental methods can be developed to discriminate between the possible locations of $V_B^-$ within the h-BN layer. If a defect is located near an edge, then how the edge is structured and terminated are additional important issues.

**Results**

**Negatively Charged Boron Vacancy**

The ground state of this defect is expected to have D$_{3h}$ local point group symmetry[9] if located deep within the h-BN layer. The ground state of this system is predicted to be $(1)^3A_2'$,[20] consistent with properties observed by Gottscholl et al.[13] for their defects displaying ODMR. (Note that we use standard group-theoretical axis conventions throughout.[26]) Our results for the excited states of this defect are summarized in Table 1. In the centre of a h-BN layer, its first triplet excited state is predicted to be $(1)^3E''$, a state that undergoes a Jahn-Teller distortion to produce the $(1)^3A_2$ state in C$_{2v}$ symmetry at an adiabatic transition energy of 1.71 eV. This quantity is a good approximation for the ZPL energy and is in excellent agreement with observed[13] PL properties. The transition is formally forbidden, consistent with the very weak observed PL signal. However, the calculated reorganization energy in emission $\lambda^E$ of 0.29 eV that specifies the PL bandwidth is double the largest values that could be consistent with observations. Overestimation of reorganization energies is a known problem with the use of model compounds for defects as they do not embody enough external constrain on the nuclear motion.[10] These properties are consistent with other predictions for bulk $V_B^-$ defects;[20] in addition, Table 1 shows that only small changes to most PL properties are expected when $V_B^-$ is located near a h-BN edge, though the reorganization energies do significantly increase.

**Table 1.** CAM-B3LYP TDDFT-calculated state properties of the $V_B^-$ and $V_N^-$ defects in a 3-ring model compound with the defect located at the centre or near an edge, including: the adiabatic transition energy $\Delta E_0$ (in eV), the vertical excitation energy $E_v^A$ (in eV), and the absorption and emission reorganization energies $\lambda^A$ and $\lambda^E$, respectively (in eV).[a]

| Defect | in centre | | | | | at boron zigzag edge | | | | |
|---|---|---|---|---|---|---|---|---|---|---|
| | state | $\Delta E_0$ | $\Delta E_v^A$ | $\lambda^A$ | $\lambda^E$ | state | $\Delta E_0$ | $\Delta E_v^A$ | $\lambda^A$ | $\lambda^E$ |
| $V_B^-$ | $(1)^3A_2'$ | [0] | | | | $(1)^3B_2$ | [0] | | | |
| | $(1)^3E'' \to (1)^3A_2$[c] | 1.71 | 2.00 | 0.29 | 0.29 | $(1)^3A_2$[bc] | 1.74 | 2.13 | 0.39 | 0.53 |
| | $\to (1)^3B_1$ | 1.87[b] | 2.00 | 0.13 | 0.11 | $(1)^3B_1$ | 1.65 | 2.14 | 0.49 | 0.20 |
| | $(1)^3A_1''$ | 2.01 | 2.05 | 0.04 | 0.04 | $(2)^3A_2$[c] | 1.70 | 2.32 | 0.62 | 0.38 |
| $V_N^-$ | $(1)^1A_1'$ | [0] | | | | $(1)^3B_1$ | [0] | | | |
| | $(1)^1E'' \to (1)^1A_2$ | 2.77 | 3.01 | 0.24 | 0.82 | $(2)^3B_1$[c] | 2.22 | 2.79[d] | 0.57 | 0.56 |
| | $\to (1)^1B_1$[c] | 1.17[e] | 3.01 | 1.84 | 1.59 | $(3)^3B_1$[c] | 2.03 | 2.90 | 0.87 | 0.89 |
| | $(1)^1E' \to (3)^1A_1$[c] | | 3.65 | | | $(1)^3A_2$ | 2.32 | 2.94 | 0.62 | 0.60 |
| | $\to (1)^1B_2$[c] | | 3.65 | | | $(2)^3A_2$ | | 3.01 | | |
| | $(1)^3E'' \to (1)^3A_2$ | 0.31[ef] | 2.21 | 1.90 | 0.49 | $(1)^1A_1$ | 1.19 | 1.75 | 0.56 | 0.46 |
| | $\to (1)^3B_1$ | 0.49[ef] | 2.21 | 1.72 | 1.37 | | | | | |
| | $(1)^3E' \to (1)^3A_1$ | 1.09 | 2.62 | 1.53 | 0.49 | | | | | |
| | $\to (1)^3B_2$ | 1.95 | 2.62 | 0.67 | 0.72 | | | | | |

a: Arrows indicate the symmetry lowering[26] either mandated by Jahn-Teller distortion or otherwise from D$_{3h}$ to C$_{2v}$ symmetry; $V_N^-$ in the centre is also unstable owing to warping of the plane to C$_2$ symmetry. CAM-B3LYP TDDFT calculations typically underestimate isolated-layer transition energies by 0.3±0.2 eV in situations relevant to $V_B^-$ and $V_N^-$ at the edge, increasing to 0.4±0.3 eV for $V_N^-$ at the centre;[10] no corrections are made for the embedding of the defect layer inside bulk h-BN.

b: presumably forms a transition-state on the (possibly distorted) Jahn-Teller potential-energy surface.

c: Franck-Condon allowed transitions between the ground state and this state.

d: adsorption vertical oscillator strength 0.028.

e: when $\lambda^A + \lambda^E > E_v^A$, an energetically accessible conical intersection with the ground state is present, making an ultrafast relaxation pathway for initial excitation back to the ground state as well as for relaxation to a potentially stable defect isomer.

f: lowest vertical excitations from these triplet states are of B$_1$ type, at 2.19 eV for $(1)^3A_2$ and 1.91 eV for $(1)^3B_1$.

In the $(1)^3A_2'$ ground state of $V_B^-$, two unpaired electrons occupy an $e'$ defect lone-pair type orbital of σ type, leading to a calculated ZFS of $D$ = 3.32 GHz. This agrees well in magnitude with the value of -3.6 GHz deduced by Gottscholl et al.[13] from EPR and ODMR measurements at T = 5 K, but has opposite sign, as also has been previously reported.[20] Even at low temperature, however, thermal population of the Zeeman sublevels makes determination of the sign of $D$ problematic, hence the predictions both here and elsewhere[20] may prove reliable, indicating that the ODMR observed by Gottscholl et al.[13] is associated with the $V_B^-$ defects inside h-BN layers.

We pursue the option that the sign determination by Gottscholl et al.[13] is in fact correct, considering the location of $V_B^-$ away from step edges. The geometric model used for this is shown in Fig. 1(b), with the defect located in an h-BN strip at different positions in from the step edge. A boron zigzag edge is used that is H-terminated. The calculated ZFS changes slowly as the defect approaches the edge from the bulk until the edge is reached, at which point it dramatically changes sign, as shown in Fig. 1(a). At the edge it is -3.5 GHz, in good agreement with the experimentally derived conclusion.[13] The calculated HF coupling constants for $V_B^-$ at the centre of h-BN sheet are $A$ = [49.1, 54.5, 102.5] MHz, while for the defect located at the edge are $A$ = [47.2, 49.39, 89.5] MHz, both in good agreement with measured $xy$-plane values[13] of 47 MHz.

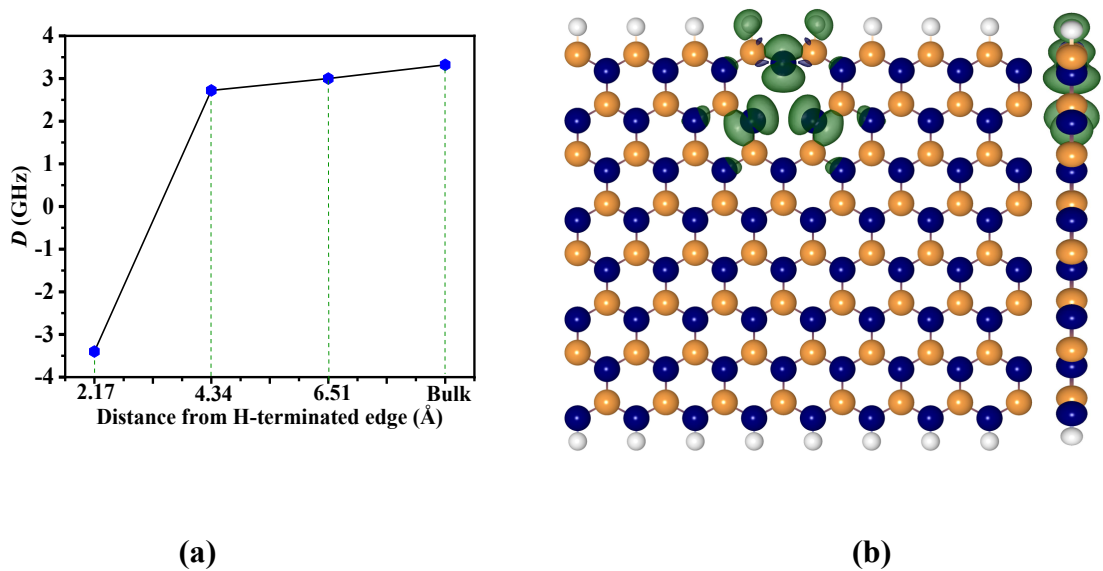

(a)            (b)

**Fig 1.** (a) The ZFS parameter $D$ calculated for $V_B^-$ as a function of the distance of the defect centre from an H-passivated edge of h-BN. (b) An isosurface of the calculated spin density

for $V_B^-$ close to H-passivated zig-zag edge of h-BN at a value 0.001/°Å$^3$. The structure shown is periodic in the *x*-direction and located in the *xy* plane. For $V_B^-$ in bulk h-BN *S* = 1, *D* = 3.32 GHz, *A* = [49.1, 54.5, 102.5] MHz and close to H-terminated edge *S* = 1, *D* = -3.5 GHz, *A* = [47.2, 49.39, 89.5] MHz.

**Negatively Charged Nitrogen Vacancy**

This defect in the bulk has been predicted to have a closed-shell singlet $(1)^1A_1'$ ground state,[19] and our calculations support this prediction. ODMR can be observed for systems with singlet ground states that allow for intersystem crossings (ISC) from initially excited singlet states to triplet states,[27] but this requires also significant buildup of population on the triplet state. The calculated PL properties of the singlet ground state are listed in Table 1 and indicate that it is highly unlikely that this could lead to any of the observed ODMR signals: these signals occur following excitation by light with under 2 eV energy per photon, whereas the lowest-energy singlet excited state is predicted to be at 3 eV. Further, the lowest excited state is predicted to undergo a huge Jahn-Teller distortion with a reorganization energy of 1.19 eV, two orders of magnitude larger than observed spectral widths. It also manifests out-of-plane buckling and a conical intersection with the ground state that would facilitate extremely fast non-radiative energy loss. Of note, however, is that our calculations predict that the lowest-energy triplet states are just 0.3 – 0.5 eV higher in energy than $(1)^1A_1'$, numbers of the order of likely errors[10] in the computational approach and hence the ground state may indeed by triplet in character. The lowest-energy transitions within the triplet manifold are predicted to be out-of-plane polarized at 2.19 eV for $(1)^3A_2$ and 1.91 eV for $(1)^3B_1$, consistent with observed PL properties.

Table 1 also shows that moving the defect to an edge has a profound effect on the properties of the ground and excited states of $V_N^-$, making a firm prediction of a $(1)^3B_1$ ground state. For the edge-located defect used in the calculations (see SI Section S3), the excitation energies within the triplet manifold now appear too large to account for the ODMR results, as also do the reorganization energies. Nevertheless, the calculations raise the possibility that some other edge-related $V_N^-$ defect could have the observed PL properties as they are established as being strongly site dependent.

For the lowest triplet state, our calculations predict a ZFS of *D* = -46 MHz for $V_N^-$ in bulk h-BN and -30 MHz close to an H-passivated edge (Fig. 2(a)), (hfc tensor *A* = [10.0, 10.0, 29.5]

MHz for bulk and *A* = [11.11, 11.08, 30.01] MHz for H-passivated edge).  This rules out the possibility of this defect also being responsible for the ODMR/EPR observations by Gottscholl et al.[13] as the |*D*| is predicted to be only one hundredth of the observed value.  Alternatively, the ODMR measurements of Chejanovsky et al.[14] give an upper bound on the value of |D| = 4 MHz.  This extremely small value is qualitatively consistent with the computational predictions for $V_N^-$, and the defect has an unpaired electron in a π orbital, as believed.[14]  Further, the observed value for the hfc is around 10 MHz is in excellent agreement with that predicted for $V_N^-$.  These results, combined with those for PL, strongly suggest that the ODMR observed by Chejanovsky et al.[14] can be attributed to $V_N^-$ located near step edges.

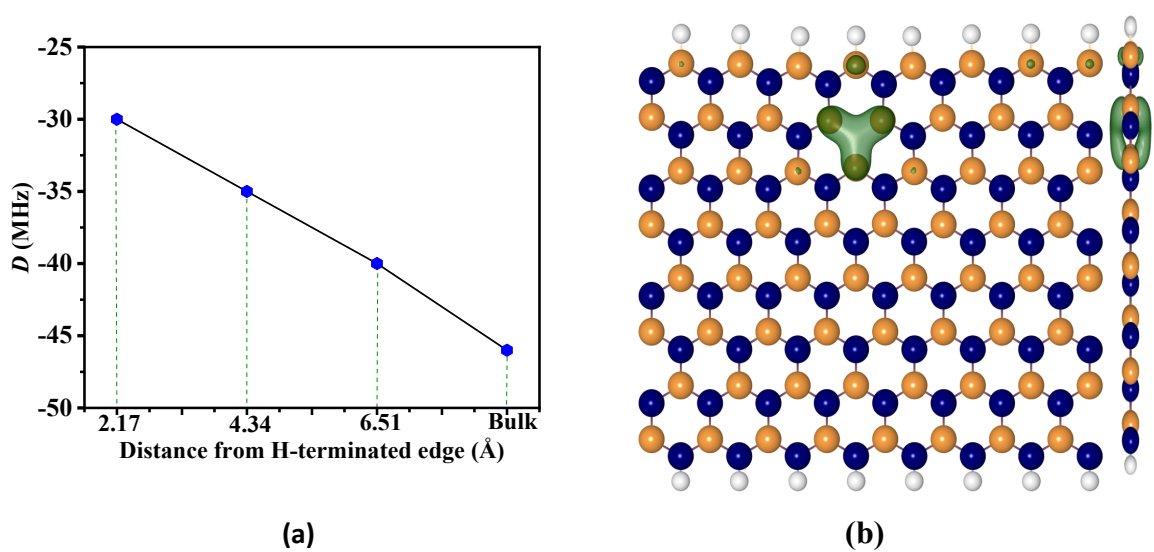

(a)     (b)

**Fig 2.** (a) ZFS parameter *D* for $V_N^-$ as a function of distance form H-passivated edge of h-BN (b)  The isosurfaces of calculated spin density for $V_N^-$ close to H-passivated zig-zag edge of h-BN at an isovalue of 0.001/°Å³. Spin density is concentrated on the atoms, providing significant hyperfine couplings. For $V_N^-$ in bulk h-BN *S* = 1, *D* = -46 MHz, *A* = [10.01, 10.01, 29.0] MHz and close to H-terminated edge *S* = 1, *D* = -30 MHz, *A* = [11.11, 11.08, 30.01] MHz.

It is no easy task to rationalize the observation of edge-located defects in which the edge is fully H-terminated yet the defect contains no hydrogen.  Sample preparation conditions would be very important to such an occurrence.  Defect production would normally be thought to be under kinetic control rather than thermodynamic control, especially for defects related away from edges. We note one property relevant to this discussion: in Fig. 3, the relative formation energies of the defects considered in periodic h-BN strips are plotted as a function of the distance of the defect from the edge. From a purely thermodynamic standpoint, edge-based defects are hence seen to be more stable than central defects.

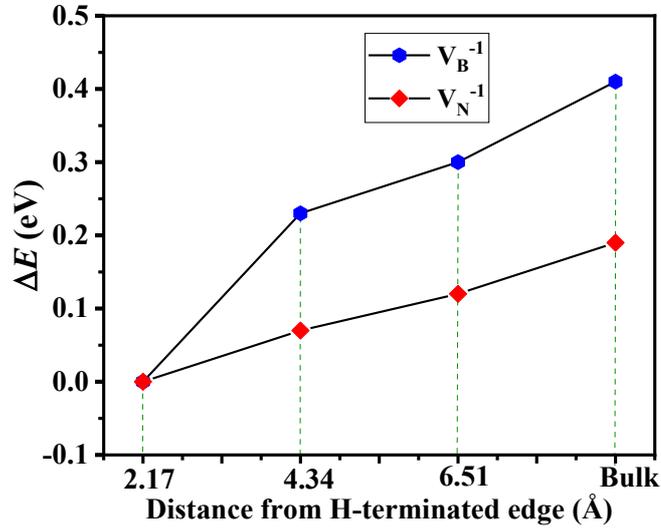

**Fig. 3.** Total energy of the $V_B^-$ and $V_N^-$ as a function of distance from H-passivated edge.

**Microscopic Origin of Changes in Zero Field Splitting**

Lattice strains are already known to influence the spectral properties of some SPEs.[28] We compute $D$ as a function of hydrostatic pressure, for both defects $V_B^-$ and $V_N^-$ in bulk, as shown in Fig. 4. This hydrostatic pressure produces isotropic compressive and tensile strains around the defect centers, thus preserving the $D_{3h}$ point group symmetry, with $D$ found to vary linearly with strain. In our calculations, the strain dependent ZFS yields a slope of 7.9 MHz/GPa (for +ve strain) for $V_B^-$. This number, in principle, can be compared to the experimentally observed variation in ZFS with temperature, which has been attributed to thermal expansion of the lattice, yielding a slope of -0.4 MHz/K, equivalent to a pressure coefficient of 2.9 MHz/GPa.[13] These results do not entirely agree, unlike good agreement found between analogous calculations and observed properties for the $N_V^{-1}$ center in diamond.[17] If the defect were located at a step edge, then thermal expansion would lead to slippage in the direction orthogonal to the edge and hence the sensitivity of the ZFS would be reduced, as observed, suggesting that the defect is located at a step edge.

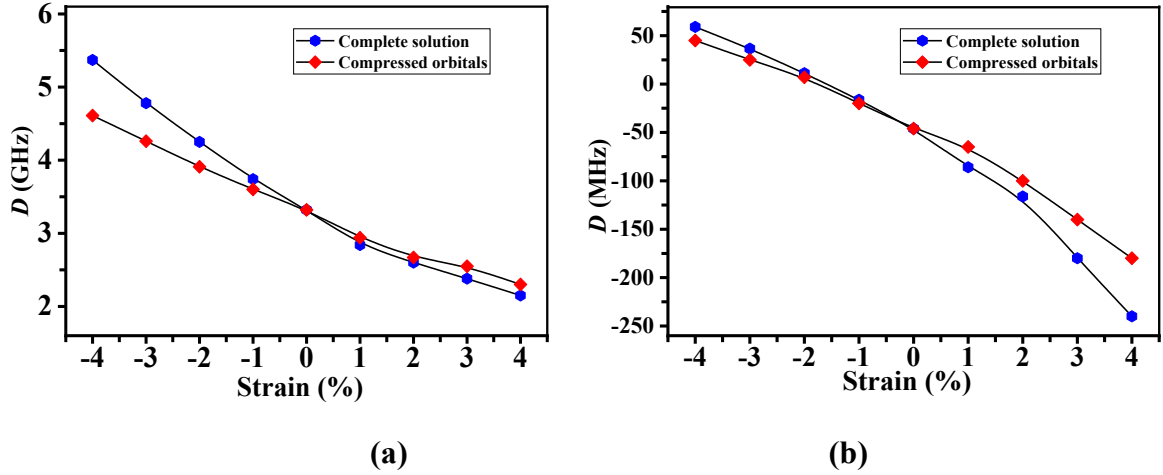

**Fig. 4.** Calculated pressure dependence of ZFS for (a) $V_B^-$ and (b) $V_N^-$, using two different models. The green line shows the results of a complete self-consistent solution (with geometry relaxation) of the strained supercells of h-BN embedding a single defect. The purple line represent the ZFS calculated according to analytical compressed orbital model described in ref.[17] In this model, only the distance of the spins is affected, while the change of the spin density distribution is neglected.

One may distinguish two contributions to the variation of ZFS as a function of pressure: purely geometrical changes around the defect center and the variation of the defect's spin density. The former may be described using the compressed-orbital model,[17] according to which the ZFS becomes scaled by a geometrical factor ($d/d_0$) determined by the atomic relaxation induced by pressure, in the proximity of the defect; here $d$ and $d_0$ are the nearest-neighbour distances under pressure and at equilibrium, respectively. As shown in Fig. 4, the compressed-orbital model describes well the calculated pressure dependence of ZFS for both $V_B^-$ and $V_N^-$, showing a negligible contribution arising from spin density changes. This is consistent with the results for $N_v^{-1}$ centre in diamond, where the change in ZFS is attributed predominantly to geometrical changes.[17, 29]

    We now discuss the microscopic origin of the calculated change in $D$ from the bulk values of 3.32 GHz in $V_B^-$ and -46 MHz in $V_N^-$ to the edge-values of -3.5 GHz and -30 MHz, respectively. As the edge is approached, the nearest-neighbour bond lengths inside the defect change from three equivalent 2.59 Å separations to one at 2.59 Å and two at 2.81 Å for $V_B^-$, as highlighted in SI Fig. S1; for $V_N^-$, these distances are 2.04 Å and 2.30 Å or 2.04 Å, respectively, embodying the lowering of local point-group symmetry from $D_{3h}$ to $C_{2v}$ as the

edge is approached. This geometrical change is responsible for the changes in ZFS parameters. The fact that the observed value of *E* is non-zero[13] may also be attributed to this effect. Our calculated value of *E* for the defect at the edge is 80 MHz, which on an absolute scale is in a reasonable agreement with the experimental value of 50 MHz.[13]

Calculations also show that the calculated ZFS parameters are very sensitive to the nature of any edge to which that may be near. Optimized geometries for the $V_B^-$ and $V_N^-$ defects adjacent to periodic zig-zag edges are also shown in Fig. S1, with isodensity surfaces akin to Figs. 1 and 2 shown in SI Figs. 2-5. No sign change is predicted at the zigzag edge for $V_B^-$ as the associated geometrical changes are only small, but the changes for $V_N^-$ are very large, suggesting that some spin redistribution is occurring.

Some other chemical structures are considered in SI. Figs. S6 and S7. First, a complex defect is considered in which many atoms are missing and the defect is internally hydrogen terminated, all except for a single remaining boron dangling bond.[11] For this, the calculated ZFS parameter is *D* = 0.41 GHz, inconsistent with all current experimental observations of defects displaying ODMR. Next, we consider the $C_N^\pm$ defects in which a positively or negatively charge carbon atom replaces a nitrogen. Again, the calculated ZFS parameter of -0.54 GHz for $C_N^-$ and -0.51 GHz for $C_N^+$, inconsistent with all known observations.

Finally, we consider the effect that embedding a defect-containing layer inside other complete h-BN layers is considered in SI Fig. S8. For a defect away from edges, the ZFS parameter is predicted to change insignificantly in response to the presence of the surrounding material.

In conclusion, we stress that the optical and magnetic properties of defects may change considerably as a function of the location of the defect in regard to step edges in h-BN. For $V_B^-$, the calculated ZFS parameter *D* changed sign on going from edge to centre, with only small changes to the PL, whereas for $V_N^-$, *D* only changed slightly for the triplet state but the nature of the ground state changed from triplet to singlet. Defect location and defect properties are strongly correlated.

The calculations presented here for ground state spin properties and the PL spectra provide conclusive evidence that the recently observed ODMR/EPR spectra for h-BN by Gottscholl et al.[13] arises from the $V_B^-$ defect. Our calculated values for the magnitude of the ZFS parameter *D* and hfc tensor *A* are in excellent agreement with those observed by Gottscholl et al.[13] We find that the sign of *D* changes when the defect is located at a hydrogen passivated boron zig-zag edge from 3.32 GHz to -3.5 GHz. The observed value is -3.6 GHz, albeit with doubt remaining over the sign. The calculated value for the hfc tensor for $V_B^-$ in bulk h-BN is *A* = [49.1, 54.5, 102.5] MHz and for $V_B^-$ close to H-passivated edge is *A* = [47.2, 49.39, 89.5] MHz compared with the measured value of 47 MHz. Furthermore, locating the defect at the edge reduces its symmetry to $C_{2v}$ and can therefore explain the observed non-zero value for the non-axial ZFS parameter *E*. The value of *E* in our calculations i.e. 80 MHz is in a reasonable agreement with experimental value of 50 MHz. In light of these results, an unambiguous measurement of the sign of *D* will therefore determine the location of the emitter.

Similarly, the ODMR observations of Chejanovsky et al.[14] for Group-2 SPEs can be tentatively assigned to $V_N^-$ defects located near step edges. At the edge, the calculations strongly suggest a triplet ground state, with a singlet ground state slightly preferred in the centre with PL properties similar to those observed. The ZFS parameter *D* varies between -46 MHz and -30 MHz for location in the bulk and at the edge respectively and hfc tensor of *A* = [10.0, 10.0, 29.5] MHz. These results are consistent with the experimental observations, whereas a simple defect involving carbon, as was originally suggested as a possible center, was found to be inconsistent. The relationship of these results to more widely obtained results for Group-2 SPE's[10] needs to be established.

**Methods**

The computational parameters and theoretical details are presented in detail in Supporting Information (SI) Section S.1.[30, 31, 32, 33, 34, 35] Briefly, periodic DFT calculations are performed by the Vienna Ab Initio Simulation Package (VASP)[30, 31] using the non-local Heyd-Scuseria-Ernzerhof hybrid functional (HSE06)[34, 35] in replicated images representing h-BN monolayers, strips, and multi-layer slabs. For the strips, we assume that all h-BN edges are hydrogen terminated, with no access of hydrogens to partially fill defect vacancy sites, even if the defect is located adjacent to an edge. Magnetic properties are evaluated using standard methods.[17, 19, 36, 37] Excited-state properties are evaluated on a 3-ring model compound (see SI Section S3) representing the defects, evaluating transition energies, reorganization energies, and oscillator strengths by TDDFT using Gaussian-16.[38] The CAM-B3LYP density functional[39, 40, 41] is used with the 6-31G* basis set,[42] this being a calibrated entry-level method for the evaluation of defect spectroscopic properties.[15]

**Acknowledgements**

This work was supported by resources provided by the National Computational Infrastructure (NCI), Pawsey Supercomputing Centre, Australia and funding from the Australian Research Council (DP150103317 and DP160101301). K.S.T. acknowledges funding from the European Research Council (ERC) under the European Union's Horizon 2020 research and innovation program (Grant Agreement No. 773122, LIMA).

**Author Contributions**

A.S conceived and proposed the main idea of this work. A.S and J.R.R contributed to all aspects of this work, with A.S contributing calculations of the spin-state properties and J.R.R the electronic state calculations. MJF contributed to editing of the manuscript and analysis of results particularly with regard to understanding the previous literature and the relevance of the current work. K.S.T contributed to discussions regarding the design and interpretation of the nano ribbons. All the Authors contributed to the analysis of results and writing of manuscript.